\documentclass{article}

\usepackage{epsfig,float}
\newcommand{\bfr}{\begin{flushright}}
\newcommand{\efr}{\end{flushright}}

\begin{document}
\title{
Double logarithms resummation in exclusive processes : the surprising behavior of DVCS
\thanks{Presented at  the Low x workshop, May 30 - June 4 2013, Rehovot and
Eilat, Israel}
}
\author{T. Altinoluk $^1$,
B. Pire $^2$, L. Szymanowski $^3$ and S. Wallon $^4$\\
{\small1.Departamento de F\'\i sica de Part\'\i culas and IGFAE,}\\
{\small Universidade de Santiago de Compostela,}\\
{\small E-15782 Santiago de Compostela, Galicia-Spain}\\
{\small 2. CPhT, \'Ecole polytechnique, CNRS, F91128 Palaiseau, France}\\
{\small 3. National Center for Nuclear Research (NCBJ), Warsaw, Poland}\\
{\small 4. LPT, Universit{\'e} Paris-Sud, CNRS, 91405, Orsay, France {\em \&} }\\
{\small UPMC Univ. Paris 06, facult\'e de physique, 4 place Jussieu,}\\
{\small 75252 Paris Cedex 05, France}
\smallskip\\
}
\date{\today
}
\maketitle
\begin{abstract}
Double logarithms resummation has been much studied in inclusive as well as  exclusive processes. The Sudakov mechanism has often be the crucial tool to exponentiate potentially large contributions to amplitudes or cross-sections near phase-space boundaries. We report on a recent work where a very different pattern emerges : the  DVCS quark coefficient function $C^q(x,\xi)$ develops near the particular point $x=\xi$  a {\em non-alternate} series in $\alpha_s^n \log^{2n}(x-\xi)$ which may be resummed in a $\cosh[K\sqrt \alpha_s \log(x-\xi)]$ factor. This result is at odds with the known result for the corresponding coefficient function for the pion transition form factor near the end point $C^q(z)$ although they are much related through a $z\rightarrow x/\xi$ correspondence.
\\
~
\\
Preprint numbers: CPhT-PC-081-0813; \hspace{.2cm} LPT-ORSAY-13-62
\end{abstract}

\section{Introduction}
While perturbative calculations are widely used in quantum field theory, their summation is always
a formidable task, unreachable but in the simplest, not to say most simplistic, occurences. From
elementary particle to atomic and to solid state physics, resummation techniques have been
developped to go beyond a fixed order perturbation estimate through the sampling and evaluation of
an infinite class of diagrams shown to dominate in a limited kinematical region for some given
observables. The result of such procedures is often an exponentiated factor as in the famous
Sudakov case~\cite{Sud}, $\exp(- K g^2 \log^2z)$, where $g$ is the coupling constant and $z$ is the ratio of two different
characteristic  scales,  which governs both QED and QCD calculations of exclusive form
factors. As explained in detail in \cite{APSW1} we obtain for a specific case of exclusive scattering amplitude, namely the deeply virtual Compton scattering in the generalized Bjorken regime,  a very different resummed result of
the form $\cosh (K g \log z) $ where $z$ is a momentum fraction. To our knowledge, this form never
previously emerged in field theoretic calculations. The process that we focus on is the most studied case of a class of reactions - exclusive hard hadronic
processes - which are under intense experimental investigation. The result presented here
 provides an important stepping-stone for further developments enabling a consistent extraction of
the quantities describing the 3-dimensional structure of the proton.

In the  collinear factorization framework the scattering amplitude for  exclusive processes such as deeply virtual Compton scattering (DVCS)
has been shown \cite{historyofDVCS} to factorize in specific kinematical regions, provided a large scale controls the separation of short distance dominated partonic subprocesses and long distance
hadronic matrix elements, the generalized parton distributions (GPDs) \cite{review}.
The  amplitude for the DVCS process
\begin{equation}
\gamma^{(*)}(q) N(p) \to \gamma(q') N'(p')\,,
\label{reaction}
\end{equation}
with a large  virtuality $q^2 =-Q^2$, factorizes in terms of perturbatively calculable coefficient functions $C(x,\xi, \alpha_s) $ and GPDs $H(x,\xi,t)$, where the scaling variable in the generalized Bjorken limit is the skewness $\xi$ defined as
$
\xi = \frac{Q^2}{(p+p')\cdot(q+q')}
$.
The calculation of first order perturbative corrections to the
partonic amplitude has shown that terms of order
$\frac{\log^2(x\pm\xi)}{x\pm\xi}$ play an important role in the region of small $(x\pm\xi)$
i.e.
in the vicinity of the boundary between the domains where the QCD
evolution equations of GPDs take distinct forms (the so-called ERBL and DGLAP domains). We scrutinize these
regions and demonstrate that they are dominated by soft fermion and
gluon propagation. This explains why they can be exponentiated using
quasi-eikonal techniques.


\section{Main steps of our analysis.
\label{sec2}}
To set up our notations, let us remind the reader  of the known results for the NLO corrections to the  DVCS amplitude  (\ref{reaction}), specializing to
 the quark contribution to its symmetric part. After proper renormalization,
it reads
\begin{eqnarray}
\mathcal{A}^{\mu\nu} = g_T^{\mu\nu}\int_{-1}^1 dx
\left[
\sum_q^{n_F} T^q(x) H^q(x,\xi,t)
\right]\,,
\label{eq:factorizedamplitude}
\end{eqnarray}
where the  quark coefficient function $T^q$ reads \cite{Pire:2011st}
\begin{eqnarray}
\!\!\!\!\!\!T^q&\!\!=&\!\! C_{0}^q +C_1^q +C_{coll}^q \log \frac{|Q^2|}{\mu^2_F}  \,,\\
\!\!\!\!\!\!C_0^q &\!\!=&\!\! e_q^2\left(\frac{1}{x-\xi+i\varepsilon} \,-\, (x \to -x)  \right) \,, \label{C0} \\
\!\!\!\!\!\!C_1^q &\!\!=&\!\! \frac{e_q^2\alpha_SC_F}{4\pi(x-\xi+i\varepsilon)}
\bigg\{
\log^2(\frac{\xi-x}{2\xi}-i\varepsilon )
\,-\,  9 \nonumber \\
&&- \, 3\frac{\xi-x}{\xi+x}\log\bigg(\frac{\xi-x}{2\xi}-i\epsilon\bigg)\bigg\}-(x \to -x) \,.
\label{C1}
\end{eqnarray}
The first (resp. second) terms in Eqs. ~(\ref{C0}) and (\ref{C1}) correspond to the $s-$channel (resp. $u-$channel) class of diagrams. One goes from the $s-$channel to the  $u-$channel by the interchange of the photon attachments. Since these two contributions are obtained from one another by   a  simple ($x\leftrightarrow -x$) interchange, we now restrict  mostly to the discussion of the former class of diagrams.

Let us first point out that in the same spirit as for evolution equations, the extraction of the soft-collinear singularities which dominate the amplitude in the limit $x \to \pm \xi$ is made easier if one uses the light-like gauge $p_1\cdot A=0$ with $p_1 = q'$. We argue (and verified) that in this gauge the amplitude is dominated by ladder-like diagrams. We expand any momentum in the Sudakov basis $p_1$, $p_2$,
as $k = \alpha \, p_1 + \beta \, p_2 + k_\perp\,,$
where $p_2$ is the light-cone direction of the two incoming and outgoing partons ($p_1^2=p_2^2=0$, $2 p_1 \cdot p_2 =s = Q^2/2\xi $).
 In this basis,  $q_{\gamma^*}=p_1-2 \, \xi \, p_2$\,.

We now restrict our study to the limit   $x \to +\xi$. The dominant kinematics is given by a strong ordering both in longitudinal and transverse momenta, according to (see Figure 1) :
\begin{eqnarray}
&& \hspace{-.3cm}x \sim \xi\gg \vert \beta_1\vert  \sim \vert x-\xi \vert \gg \vert x-\xi +\beta_1\vert \sim \vert \beta_2 \vert \gg \cdots \nonumber \\
&&\hspace{-.4cm}
\cdots \gg   \vert x-\xi +\beta_1 +\beta_2 +\cdots+ \beta_{n-1} \vert  \sim  \vert \beta_n  \vert  ,
\label{kinematics_beta}
\\
&& \vert k_{\perp 1}^2\vert  \ll \vert k_{\perp 2}^2\vert  \ll \cdots \ll \vert k_{\perp n}^2\vert  \ll s \sim Q^2 \,,
\label{kinematics_k}
\\
&&  \vert \alpha_1 \vert  \ll \cdots \ll  \vert \alpha_n \vert  \ll 1\,.
\label{kinematics_alpha}
\end{eqnarray}
This ordering is related to the fact that the dominant double logarithmic contribution for each loop arises from the region of phase space where both soft and collinear singularities manifest themselves. When $x \to \xi$ the left fermionic line is a hard line, from which the gluons are emitted in an eikonal way (which means that these gluons have their all four-components neglected in the vertex w.r.t. the momentum of the emitter), with an ordering in $p_2$ direction and a collinear ordering. For the right fermionic line, eikonal approximation is not valid, since the dominant momentum flow along $p_2$ is from  gluon to  fermion, nevertheless the collinear approximation can still be applied.

When computing the coefficient functions, one faces both UV and IR divergencies. On the one hand, the UV divergencies are taken care of through renormalization, which manifest themselves by a renormalization scale $\mu_R$  dependency. On the other hand, the IR divergencies remain, but factorization proofs at any order for DVCS justify the fact that they can be absorbed inside the generalized parton distributions and result in finite coefficient functions.
In our study, we are only interested into finite parts. Thus, using dimensional regularization, in a factorization scheme like $\overline{MS}$, any scaleless integral can be safely put to zero  although it contains both UV and IR divergencies. Following this line of thought, we can thus safely deal with DVCS on a quark for our resummation purpose.

Finally,
the issue related to the $i \epsilon$ prescription in Eq.~(\ref{C1}) is solved by computing the coefficient function in the unphysical region $\xi
> 1$. After analytical continuation to the physical region $0 \leq \xi \leq 1$, the final result is then obtained through the shift $\xi \to \xi-i \epsilon$\,.

We define  $K_n$ as the contribution of a $n$-loop ladder to the coefficient  function.
Let us sketch the main steps of the derivation of $K_1$ and then generalize it for $K_n$.

 \begin{figure}
  \epsfig{figure= 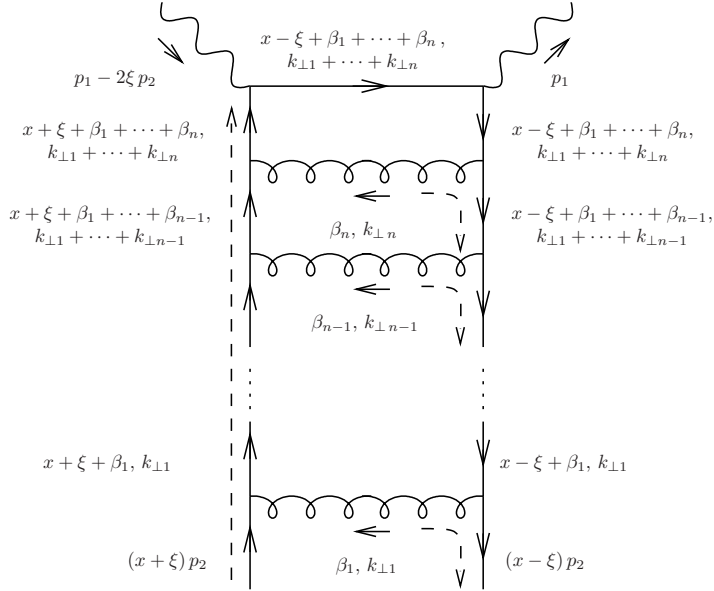 , height=8cm}
 \caption{\small The ladder diagrams which contribute in the light-like gauge to the leading  $\alpha_s^n \ln^{2 n}(\xi-x)/(x-\xi)$ terms in the perturbative expansion of the    DVCS  amplitude. The $p_2$ and $\perp$ momentum components are indicated. The dashed lines show the dominant momentum flows along the $p_2$ direction.} 
 \label{Fig:n-loop}
 \end{figure}

\paragraph*{The ladder diagram at order $\alpha_s$.}
A careful analysis \cite{APSW1}  shows that among the one loop diagrams and in the light-like axial gauge $p_1.A=0$, the box diagram is dominant for $x \to \xi\,$.
Starting from the dominant part of the numerator of the Born term which is $\not{\theta}=-2\not{p}_1$,
the numerator of the box diagram is
\begin{equation}
tr\big\{ \not{p_2}\gamma^{\mu}[\not{k}+(x-\xi)\not{p_2}]\not{\theta}[\not{k}+(x+\xi)\not{p_2}]\gamma^{\nu} \big\}d^{\mu\nu}\,.
\end{equation}
In the limit $x\to\xi$, while the left fermionic line is hard with a large $p_2$ momentum, the gluonic line is soft with respect to the left fermionic line. So we perform soft gluon approximation in the numerator by taking  $k+(x+\xi)p_2 \to (x+\xi)p_2$. The dominant contribution comes from the residue of the gluonic propagator. Thus, the numerator of the on-shell gluon propagator, $d^{\mu\nu}$, is expressed in terms of transverse polarizations, i.e. $
d^{\mu\nu}\approx -\sum_{\lambda}\epsilon^{\mu}_{(\lambda)}\epsilon^{\nu}_{(\lambda)}\,.
$

Writing the gluon polarization vectors in the light-like $p_1\cdot A=0$ gauge through their Sudakov decomposition
\begin{equation}
\epsilon^{\mu}_{(\lambda)}= \epsilon^{\mu}_{\perp (\lambda)} -2 \frac{\epsilon_{\perp (\lambda)}\cdot k_{\perp}}{\beta s}p_1^\mu \,,
\end{equation}
allows us to define  an effective vertex for the gluon and outgoing quark through the polarization sum
\begin{equation}
\sum_{\lambda}\epsilon_{\perp (\lambda)}\cdot k_{\perp}\epsilon^{\mu}_{(\lambda)}=-k_{\perp}^ \mu+2\frac{k_{\perp}^2}{\beta s}p_1^\mu \,.
\end{equation}
 The numerator, $(Num)_1$, is $\alpha-$independent and reads
\begin{eqnarray}
&&\hspace{-.7cm}\frac{-4(x+\xi)}{\beta}tr\bigg\{\!\!\not{p_2}\bigg(\not{k_{\perp}}-2\frac{k_{\perp}^2}{\beta s}\not{p_1}\bigg)[\not{k}+(x-\xi)\not{p_2}]\not{p_1}\!\!\bigg\} \nonumber \\
&=&-4(x+\xi)s\frac{2k_{\perp}^2}{\beta}\bigg[1+\frac{2(x-\xi)}{\beta}\bigg] \,.
\end{eqnarray}

We now calculate the  integral over the gluon momentum $k$,
using  dimensional regularization
$  \int d^dk \rightarrow \frac{s}{2}\int d\alpha \,d\beta \,d^{d-2}\underline{k}$, $(k_{\perp}^2=-\underline{k}^2)$.
The Cauchy integration of the gluonic pole which gives the dominant contribution reads
   \begin{equation}
 - 2  \pi i \frac{s}{2} \int_0^{\xi-x}\frac{d\beta}{s\beta }\int_0^{\infty}d^{d-2}\underline{k} \frac{(Num)_{1}}{L_1^2 \,R_1^2\, S^2}\Big{\vert}_{\alpha=\frac{\underline{k}^2}{\beta s}}
\label{oneloop}
 \end{equation}
with the denominators $L_1^2=-\underline{k}^2+\alpha(\beta+x+\xi)s$, $R_1^2=-\underline{k}^2+\alpha(\beta+x-\xi)s$, $S^2=-\underline{k}^2+(\beta+x-\xi)s$ and $k^2=-\underline{k}^2+\alpha\beta s$.
The relevant region of integration corresponds to small $\vert \beta+x-\xi \vert $. The $\beta$ and $\underline{k}$ integrations  results in our final one-loop expression :
\begin{equation}
\label{I1}
K_1= \frac{i}4 e_q^2 \left(-i \, C_F \, \alpha_s \frac{1}{(2 \pi)^2} \right) \frac{4}{x-\xi}\frac{2  \pi i}{2!}\log^2(a(x-\xi)) \,,
\end{equation}
where we kept only the most singular terms in the $x\to \xi$ region and have no control of the value of $a$ within our approximation. To fix $a$, we match our approximated one-loop result with the
full one-loop result (\ref{C1}). This amounts to cut the $\underline{k}^2$ integral at $Q^2$. The $i \epsilon$ term is included  according to the same matching. This leads to
\begin{equation}
\label{I1-final}
K_1= \frac{i}4 e_q^2 \left(-i \, C_F \, \alpha_s \frac{1}{(2 \pi)^2} \right) \frac{4}{x-\xi+ i \epsilon}\frac{2  \pi i}{2!}\log^2\left(\frac{\xi- x}{2\xi}-i \epsilon\right) \,,
\end{equation}
which is the known result. This is a positive test of the validity of our approximation procedure that we now generalize  to the $n$-rung ladder.
\paragraph*{The ladder diagram at order $\alpha_s^n$.} Let us now turn to the estimation of all $\log^{2n} (x-\xi)$ terms in the diagram shown on Fig. 1.
Assuming the strong ordering (\ref{kinematics_k}, \ref{kinematics_alpha}) in $k_{\perp}$ and $\alpha$,
the distribution of the poles generates  nested integrals in $\beta_i$ as :
\begin{equation}
\int_0^{\xi-x}d\beta_1\int_0^{\xi-x-\beta_1}d\beta_2\cdots\int_0^{\xi-x-\beta_1-\cdots -\beta_{n-1}}d\beta_n \,.
\end{equation}
The numerator for the $n^{th}$ order box diagram  is  obtained as:
\begin{eqnarray}
&&\hspace{-.4cm}(Num)_n=-4s(x+\xi)^n\frac{2k_{\perp1}^2}{\beta_1}\bigg[1+\frac{2(x-\xi)}{\beta_1}\bigg] \frac{2k_{\perp 2}^2}{\beta_2}\\
&&\hspace{-.5cm}\bigg[1\!+\!\frac{2(\beta_1+x-\xi)}{\beta_2}\bigg] \!\!\cdots \!\frac{2k_{\perp n}^2}{\beta_n}\!\bigg[1\!+\!\frac{2(\beta_{n-1}+\!\cdots \!+\!\beta_1\!+\!x\!-\!\xi)}{\beta_n}\bigg] \nonumber\!,
\end{eqnarray}
and the denominators of  propagators are, for $i = 1 \cdots n$,
\begin{eqnarray}
L_i^2&=&\alpha_i(x+\xi)s \nonumber \\
R_i^2&=&-\underline{k}_{i}^2+\alpha_i(\beta_1+\cdots +\beta_i+x-\xi)s\,, \nonumber \\
S^2& =& -\underline{k}_{n}^2+(\beta_1+\cdots +\beta_n+x-\xi)s\,.
\end{eqnarray}
Using dimensional regularization and omitting scaleless integrals, the integral   reads:
\begin{eqnarray}
&&\hspace{-2cm}\int_0^{\xi-x}\hspace{-.6cm}d\beta_1\hspace{-.1cm}\cdots\hspace{-.1cm}\int_0^{\xi-x-\cdots-\beta_{n-1}}\hspace{-.7cm}d\beta_n \hspace{-.1cm} \int_0^{\infty}\hspace{-.4cm}d^{d-2} \underline{k}_n\hspace{-.05cm}\cdots\hspace{-.15cm}\int_0^{\underline{k}_2^2}\hspace{-.3cm}d^{d-2}\underline{k}_1 (-1)^n \\
&&
\times\frac{4\, s(2\pi i)^n}{x-\xi}\frac{1}{\beta_1+x-\xi}\cdots\frac{1}{\beta_1\!+\!\cdots+\!\beta_{n-1}\!+\!x\!-\!\xi}
\nonumber \\
&&\hspace{2cm}\times
 \frac{1}{\underline{k}_1^2}\cdots\frac{1}{\underline{k}_n^2}\frac{1}{\underline{k}_n^2-(\beta_1+\cdots+\beta_n+x-\xi)s}\,.\nonumber
\end{eqnarray}
The integrals over $\underline{k}_1 \cdots \underline{k}_n$  are performed
similarly
 as in the one-loop case, resulting in:
\begin{equation}
\label{In-final}
K_n=\frac{i}4 e_q^2 \left(-i \, C_F \, \alpha_s \frac{1}{(2 \pi)^2} \right)^n\frac{4}{x-\xi+i\epsilon}\frac{(2\pi i)^n}{(2n)!}\log^{2n}\left(\frac{\xi-x}{2\xi}-i\epsilon\right) \,,
\end{equation}
where the matching condition introduced in one-loop case is extended to $n-$loops.
\section{The resummed formula.}
Based on the results Eqs.~(\ref{I1-final}, \ref{In-final}), one can build the resummed formula for  the complete amplitude; we get with $D= \sqrt{\frac{\alpha_s C_F}{2\pi}}$
\begin{eqnarray}
&&\sum_{n=0}^\infty K_n
 =
\frac{e_q^2}{x-\xi + i \epsilon} \cosh
\left[ D \log \left( \frac{\xi -x}{2 \xi } - i \epsilon \right) \right]
 \\
&&= \frac{1}2 \frac{e_q^2}{x-\xi + i \epsilon}\left[ \left( \frac{\xi -x}{2 \xi } - i \epsilon \right)^D + \left( \frac{\xi -x}{2 \xi } - i \epsilon \right)^{-D} \right] \,.\nonumber
\label{Sumn-final}
\end{eqnarray}
In the absence of a next to leading logarithmic calculation, the minimal and most natural resummed formula which has the same $O(\alpha_s)$ expression as the full NLO result, reads,  :
\begin{eqnarray}
&&\hspace{-.4 cm}(C_0+C_1)^{res}=\frac{e_q^2}{x-\xi+i\epsilon}\bigg\{
\cosh\bigg[D\log\bigg(\frac{\xi-x}{2\xi}-i\epsilon\bigg)\bigg]\nonumber \\
&&\hspace{-.4 cm}-\frac{D^2}{2}\bigg[9+3\frac{\xi-x}{x+\xi}\log\bigg(\frac{\xi-x}{2\xi}-i\epsilon \bigg)\bigg]\bigg\}-(x\rightarrow -x)\,.
\label{Res1}
\end{eqnarray}

\section{The gluon coefficient function
\label{sec5}}
For several decades the effects of gluons on many high energy processes has been widely studied. Specifically the theory of "Color Glass Condensate" shows that at very high energies the behavior of the scattering amplitudes  are dominated by gluons \cite{jimwlk}. Recently it was also shown that even at moderate energies, there are significant $O(\alpha_s)$ corrections to scattering amplitudes due to gluonic contributions for spacelike and timelike virtual Compton scatterings \cite{MPSSW}. With the above mentioned motivations performing a similar resummation procedure for gluon coefficient function of  DVCS and TCS would result in a more trustful extraction gluon GPDs.

\section{Summary and outlook
\label{sec6}}

We have demonstrated  that resummation of soft-collinear gluon radiation effects can be performed in hard exclusive  reactions amplitudes. The resulting formula for coefficient function stabilizes the perturbative expansion, which is crucial for a trustful extraction of GPDs from experimental data.
A related expression should emerge in various  reactions, such as the crossed case of timelike Compton scattering~\cite{MPSW} and exclusive meson electroproduction.

Giving these results, a question should be raised : what is the physics beyond this result, or in other words, why is the Sudakov resummation \cite{sudex} familiar to experts of hard exclusive processes not applicable here ? An even more precise question may be : how is our analysis compatible with the discussion of soft effects in the pion transition form factor, a quantity which has been much discussed \cite{stefanis} recently thanks to the experimental results of BABAR and BELLE? Let us stress that the coefficient function of this quantity is identical to the ERBL part of the coefficient function of the DVCS amplitude after a rescaling $z \rightarrow x/\xi$. Our result thus may be applied to the transition form factor. In Ref. \cite{Musatov}, it has been argued that the $\alpha_s \log^2(1-z) $ factor in the one loop expression of the coefficient function had to be understood as the sum of two very distinct terms, one of them exponentiating in a Sudakov form factor. To advocate this fact, the authors allow themselves  an excursion outside the colinear factorization framework and use the familiar detour into the coordinate space framework.  Our procedure is different and we resum the complete one loop result. In other words, one may ask to the authors of Ref. \cite{Musatov} : what happens to the remnant term proportional to  $\alpha_s \log^2(1-z) $? If indeed the usual resummation procedure of the transition pion form factor must be revised following our new results, one may ask whether the understanding of the meson form factor \cite{LiSter} should also be reconsidered.

\section*{Acknowledgments}
We thank the organisors and the French CEA (IPhT and DSM) for support. This work is  supported by  the
P2IO consortium, the Polish Grant NCN No. DEC-2011/01/B/ST2/03915, the French grant  ANR
PARTONS (ANR-12-MONU-0008-01), the Joint
Research Activity "Study of Strongly Interacting Matter"
(HadronPhysics3, Grant Agreement n.283286) under the
7$^{th}$ Framework Programm of the European Community, the European Research Council grant HotLHC ERC-2001-
StG-279579, Ministerio de Ciencia e Innovac\'\i on of Spain grants FPA2009-06867-E, Consolider-Ingenio 2010 CPAN CSD2007-00042 and FEDER.



\bibliographystyle{apsrev4-1}


\end{document}